\documentclass[prb,aps, onecolumn, showpacs]{revtex4}
\usepackage{amssymb}
\usepackage{amsmath}
\usepackage[dvips]{graphicx}
\usepackage[english]{babel}

\begin{document}

\title{"Probabilistic" approach to Richardson equations}
\author{W. V. Pogosov}
\affiliation{Institute for Theoretical and Applied
Electrodynamics, Russian Academy of Sciences, Izhorskaya 13,
125412 Moscow, Russia}

\begin{abstract}
It is known that solutions of Richardson equations can be
represented as stationary points of the "energy" of classical free
charges on the plane. We suggest to consider "probabilities" of
the system of charges to occupy certain states in the
configurational space at the effective temperature given by the
interaction constant, which goes to zero in the thermodynamical
limit. It is quite remarkable that the expression of "probability"
has similarities with the square of Laughlin wave function. Next,
we introduce the "partition function", from which the ground state
energy of the initial quantum-mechanical system can be determined.
The "partition function" is given by a multidimensional integral,
which is similar to Selberg integrals appearing in conformal field
theory and random-matrix models. As a first application of this
approach, we consider a system with the constant density of energy
states at arbitrary filling of the energy interval, where
potential acts. In this case, the "partition function" is rather
easily evaluated using properties of the Vandermonde matrix. Our
approach thus yields quite simple and short way to find the ground
state energy, which is shown to be described by a single
expression all over from the dilute to the dense regime of pairs.
It also provides additional insights into the physics of
Cooper-paired states.
\end{abstract}

\author{}
\maketitle
\date{\today }

\section{Introduction}

Bardeen-Cooper-Schrieffer (BCS) theory of
superconductivity\cite{BCS} is based on the reduced interaction
potential, which couples only electrons with opposite spins and
zero total momentum, while the interaction amplitude is taken to
be momentum-independent. It is known for a long time that
Hamiltonians of this kind are exactly solvable \cite%
{Rich1,Rich2,Gaudin,Rich3}. Namely, they lead to Richardson
equations, which are nonlinear algebraic equations for the set of
complex numbers $R_{j}$($j=1$,..., $N$), where $N$ is the number
of Cooper pairs in the system; the energy of the system is given
by the sum of all $R$'s. Originally, Richardson equations have
been derived by solving directly the Schr\"{o}dinger equation.
However, they can be also recovered through an algebraic
Bethe-ansatz approach \cite{Pogosyan}. The resolution of the
system of Richardson equations, in general case, is a hard task
\cite{Dukel}. These equations are now intensively used to study
numerically superconducting state in nanosized superconducting
systems \cite{Dukel,Braun,Gladilin}. They are also applied in the
nuclear physics, see e.g. Ref. \cite{Dukel1}

Recently, Richardson equations have been used\cite{We} to explore
connections between two famous problems: the one-pair problem solved by Cooper%
\cite{Cooper} and the many-pair BCS theory of
superconductivity\cite{BCS}. The essential ingredients of the
Cooper model are the Fermi sea of noninteracting electrons and the
layer above their Fermi energy, where an attraction between
electrons with up and down spins acts. Two additional electrons
are placed into this layer. It is also assumed that the energy
density of one-electronic states within this potential layer,
which has a width equal
to the Debye frequency, is constant. Cooper then was able to solve the Schr%
\"{o}dinger equation for two electrons. He found that the
attraction, no matter how weak, leads to the appearance of the
bound state. Cooper model was an important step towards a
microscopic understanding of superconductivity. In contrast, BCS
theory of superconductivity considers a many-pair configuration,
which includes the potential layer half-filled. Traditional
methods to tackle this problem is either to use a variational
approach for the wave
function\cite{BCS} or to apply Bogoliubov canonical transformations\cite%
{Bogoliubov}. In Ref. [\cite{We}], an \textit{arbitrary} filling
of the potential layer has been considered. Obviously, by
increasing a filling, one can attain a many-particle configuration
starting from the one-pair problem. Although such a procedure
seems to be unrealistic from the view point of current
experimental facilities, it allows one to deeper understand the
role of the Pauli exclusion principle in Cooper-paired states, as
well as to analyze a correspondence between the single-pair
problem and BCS condensate. This gedanken experiment can be also
considered as a toy model for the density-induced crossover
between individual fermionic molecules, which correspond to Cooper
problem, and strongly overlapping pairs in BCS
configuration\cite{Eagles,Leggett,Strinati}.

In order to find the ground state energy for $N$ pairs, in Ref.
[\cite{We}], a new method for the analytical solution of
Richardson equations in the thermodynamical limit was proposed.
Rigorously speaking, this method is applicable to the dilute
regime of pairs only, since it assumes that all Richardson
energy-like quantities are close to the single-pair energy found
by Cooper. By keeping only the lowest relevant correction to this
energy, the expression of the ground state energy was found to
read as
\begin{equation}
E_{N}=2N\varepsilon _{F_{0}}+\frac{N(N-1)}{\rho }-2N\left( \Omega -\frac{N-1%
}{\rho }\right) \frac{\sigma }{1-\sigma },  \label{1}
\end{equation}%
where $\varepsilon _{F_{0}}$ is the Fermi energy of noninteracting
electrons (or a lower cutoff), $\Omega $ is the potential layer
width (which corresponds to Debye frequency in Cooper model),
$\sigma =\exp (-2/\rho V)$, $\rho $ being a constant density of
energy states and $V$ an interaction amplitude. An analysis of
this equation shows that first two terms in its right-hand side
(RHS) correspond just to the bare kinetic energy increase due to
$N$ pairs, while the third term gives the condensation energy.
This condensation energy has a quite remarkable structure: it is
proportional both to the number of pairs $N$ and to the number of
empty states $\rho \Omega +1-N$ in the potential layer. Moreover,
in the limit $N\rightarrow 1$, the condensation energy per pair is
exactly equal to the binding energy of a single pair, found by
Cooper\cite{Cooper}. Thus, a pair "binding energy" in a many-pair
configuration appears to be simply a fraction of the single-pair
binding energy, the reduction being \textit{linear} in $N$ and
proportional to the number of occupied states.

Although Eq. (1) was derived for the dilute regime of pairs, it is
in a full agreement with the mean-field BCS result for the ground
state energy (originally derived for the half-filling
configuration). BCS approach has been also applied to the case of
arbitrary filling of the potential layer\cite{MWO,We}, this
analysis being similar to BCS-like models developed in Refs.
\cite{Eagles,Leggett}. The result for an arbitrary filling was
also shown\cite{MWO,We} to be consistent with Eq. (1). First two
terms in the RHS of Eq. (1) must be identified with the normal
state energy, as it appears in BCS theory. The third term is
exactly the condensation energy, apart from the fact that in BCS
theory it is traditionally written in terms of the gap $\Delta$
that masks a link with the single-pair binding energy. A detailed
discussion of these issues can be found in Refs. [\cite{We,MWO}].

In the recent paper [\cite{EPJB}], the dilute-regime approach to
the solution of Richardson equations\cite{We} has been advanced to
take into account the next-order term of the expansion, which is
needed in order to get rid of the dilute regime. It was found that
the corresponding contribution to the ground state energy is
underextensive, i.e., negligible in the thermodynamical limit.
This means that Eq. (1) derived in the dilute regime of pairs is
most likely to be universal, which implies that mean-field BCS
results are exact in the thermodynamical limit, in agreement with
earlier conclusions\cite{Bogoliubov1,Bardeen,Lieb}. Finally, very
recently the dilute-limit procedure has been extended by taking
into account all the terms in the expansion\cite{Monique}. It was
concluded\cite{Monique} that corrections beyond the dilute-limit
result of Ref. [\cite{We}] are indeed underextensive.

The aim of the present paper is to find a ground state energy
using Richardson equations without utilizing any asymptotic
expansions from the single-pair configuration and to address a
validity of Eq. (1) (beyond both a mean-field approximation and
dilute-limit expansions of Richardson equations). For that, we
suggest another way to evaluate Richardson equations analytically
in the thermodynamical limit. We
start with the so-called exact electrostatic mapping: It was noted [\cite%
{Rich3}] that Richardson equations can be obtained from the
condition of a stationarity for the "energy" of the
two-dimensional system of charges, logarithmically interacting
with each other, as well as with the homogeneous external electric
field. In the present work, we suggest to make one more step and
to consider "probabilities" to find a system of charges in a
certain configuration at the effective temperature, which is just
equal to the interaction constant $V$. This constant goes to zero
in the large-sample limit, such that the "energy" landscape is
very sharp in the vicinity of its stationary points. The
logarithmic character of the particle-particle interaction energy
in two dimensions together with our choice for the effective
"temperature" leads to the rather compact expression of
"probability". Note that it has similarities with the square of
Laughlin wave function appearing
in the theory of fractional quantum Hall effect\cite%
{Laughlin}. Next, we introduce a "partition function" and identify
the ground state energy of the initial quantum-mechanical system
with the logarithmic derivative of this function with respect to
the inverse effective temperature. This "partition function" is
actually represented by a multidimensional coupled integral.
Coupling between integration variables turns out to be similar
with that for Selberg integrals (Coulomb integrals) appearing in
conformal field theory (see, e.g., Ref. [\cite{Fateev}]), in
certain random matrix models (Dyson gas\cite{Dyson}), and in
growth problems\cite{growth}. At the same time, the structure of
the integrand for each variable indicates that it can be also
classified as the N\"{o}rlund-Rice integral.

A connection between Richardson equations and conformal field theory
has already been addressed in Ref. [\cite{CFT}], where it was shown
that BCS model can be considered as a limiting case of the
Chern-Simons theory. Indeed, it is easily seen that the structure of
Richardson equations has similarities with the structure of
Knizhnik-Zamolodchikov equations\cite{KZ}. At the same time,
Chern-Simons theory also plays a very important role in the
description of the quantum Hall effect [\cite{Girvin}]. That is why
there are similarities between the expression of "probability", as
introduced in the present paper, and Laughlin wave function. Note
that conformal field theories attract a lot of attention, since it is
assumed that they can provide a unification between quantum mechanics
and the theory of gravitation (anti de Sitter/conformal field theory
correspondence).

In the present paper, we restrict ourselves to the case of a
constant energy density of one-electronic states. We found an
efficient way to calculate the "partition function" by
transforming the integral to the sum of binomial type. Such a
transformation is possible due to the N\"{o}rlund-Rice structure
of the integrand. The sum is evaluated using properties of the
determinant of the Vandermonde matrix, this determinant being
responsible for the coupling between summation variables. We then
are able to obtain Eq. (1) and to prove its validity. We also
suggest a "probabilistic" qualitative understanding of this
result, which is related to the factorizable form of
"probability". We found that the "probability" for the system of
pairs, feeling each other through the Pauli exclusion principle,
can be represented as a linear combination of products of
"probabilities" for single pairs, each pair being in its own
environment with the part of the one-electronic levels absent.
These missing levels form two bands in the bottom as well as at
the top of the potential layer for each pair, such that the sum of
energies of single pairs for each term of a factorized
"probability" is the same.

Another method to evaluate the multidimensional integral, used in
the present paper, is to integrate it through the saddle point
corresponding to the single pair. This approach is actually based
on a well-known trick used to compute binomial sums by
transforming them to N\"{o}rlund-Rice integrals, which can be
tackled by a saddle-point method\cite{Sedgewick}. In our case,
this procedure appears to be quite similar to the solution of
Richardson equations in the dilute regime of pairs, as done in
Ref. [\cite{We}]. It gives rise to the expansion of energy density
in powers of pairs density. However, we were unable to perform
calculations along this line beyond few initial terms due to the
increasing technical complexity of the procedure. Nevertheless, we
show that only the first correction to the energy of
noninteracting pairs is extensive, while two others are
underextensive, this condition being necessary for Eq. (1) to stay
valid. Thus, the first method, based on manipulations with
binomial sums, turns out to be much more powerful and, moreover,
technically simpler. We believe that the method, suggested in the
present paper, is applicable to situations with nonconstant
density of energy states, as well as to other integrable pairing
Hamiltonians, for which electrostatic analogy
exists\cite{Amico,Ibanez}.

We also note that there already exists a method for the analytical
solution
of Richardson equations in the thermodynamical limit \cite%
{Rich3,Roman,Altshuler}. For the case of constant density of
states, this method, up to now, has been applied to half-filled
configuration only, for which its results agree with BCS theory.
In Ref. \cite{Duk2005}, it was also used for any pair density and
for the dispersion of a three-dimensional system. We here consider
arbitrary fillings at constant density of states. In addition, the
method of Ref. [\cite{Rich3}] assumes that energy-like quantities,
in the ground state, are organized into a one-dimensional
structure in the complex plane. This assumption is based on the
results of numerical solutions of Richardson
equations\cite{Rich3}. In this paper, which is purely analytical,
we would like to avoid using this assumption, that is why we have
constructed another method, which, moreover, is technically simple
and provides some additional insights to the physics of Cooper
pairs. Nevertheless, results of both approaches for the ground
state energy do coincide with each other, as well as with results
of the mean-field BCS theory.

This paper is organized as follows.

In Section II, we formulate our problem and we introduce a basis
of our "probabilistic" approach.

In Section III, we find the ground state energy in a rather simple
way, by using a representation of the "partition function" through
the coupled binomial sum. We also suggest "probabilistic"
interpretation of the obtained result.

In Section IV, we tackle integral entering the "partition
function" by integration through a single-pair saddle point and we
also establish a connection with the approach of Refs.
[\cite{We,EPJB}].

We conclude in Section V.

\section{General formulation}

\subsection{Hamiltonian}

We consider a system of fermions with up and down spins. They
attract each other through the usual BCS reduced potential,
coupling only fermions with zero total momenta as
\begin{equation}
\mathcal{V}=-V\sum_{\mathbf{k},\mathbf{k}^{\prime
}}a_{\mathbf{k}^{\prime
}\uparrow }^{\dagger }a_{-\mathbf{k}^{\prime }\downarrow }^{\dagger }a_{-%
\mathbf{k}\downarrow }a_{\mathbf{k}\uparrow }.  \label{BCSpot}
\end{equation}%
The total Hamiltonian reads as $H=H_{0}+\mathcal{V}$, where
\begin{equation}
H_{0}=\sum_{\mathbf{k}}\varepsilon _{\mathbf{k}}\left(
a_{\mathbf{k}\uparrow
}^{\dagger }a_{\mathbf{k}\uparrow }+a_{\mathbf{k}\downarrow }^{\dagger }a_{%
\mathbf{k}\downarrow }\right).  \label{H0}
\end{equation}

It is postulated that the potential $\mathcal{V}$ acts only for
the states with
kinetic energies $\varepsilon _{\mathbf{k}}$ and $\varepsilon _{\mathbf{k}%
^{\prime }}$ located in the energy shell between $\varepsilon
_{F_{0}}$\ and
\ $\varepsilon _{F_{0}}+\Omega $. In BCS theory, the lower cutoff $%
\varepsilon _{F_{0}}$ corresponds to the Fermi energy of
noninteracting electrons, while $\Omega $ is Debye frequency. We
also assume a constant density of energy states $\rho $ inside
this layer, which is a characteristic feature of a two-dimensional
system. For a three-dimensional system, this is justified,
provided that $\Omega \ll \varepsilon _{F_{0}}$. Thus, the
total number of states with up or down spins in the potential layer is $%
N_{\Omega }\equiv \rho \Omega $, while these states are located
equidistantly, such that $\varepsilon _{\mathbf{k}}=\varepsilon _{F_{0}}+$ $%
\xi _{\mathbf{k}}$, where $\xi _{\mathbf{k}}$ runs over $0$, $1/\rho $, $%
2/\rho $, ..., $\Omega $. Energy layer accommodates $N<N_{\Omega
}$ electrons with up spins and the same number of electrons with
down spins. These electrons do interact via the potential given by
Eq. (\ref{BCSpot}).

In this paper, we restrict ourselves to the thermodynamical limit, i.e., to $%
\Lambda \rightarrow \infty $, where $\Lambda $ is the system
volume. In this case, $\rho \sim \Lambda $ and $V$ $\sim \Lambda
^{-1}$, so that the dimensionless interaction constant, defined as
$v=$ $\rho V$, is volume-independent, $\sim \Lambda ^{0}$. The
same is valid for $\varepsilon _{F_{0}}$ and $\Omega $; hence
$N_{\Omega }\sim \Lambda $. The number of Cooper pairs scales as
$N\sim \Lambda $; consequently, filling $N/N_{\Omega } $ is
volume-independent, $\ \sim \Lambda ^{0}$. We treat arbitrary
fillings of the potential layer $N/N_{\Omega }$, while the
traditional BCS theory deals with the half-filling configuration.
Studies of arbitrary fillings help one to reveal an important
underlying physics, which is not easy to see when concentrating on
the half-filling configuration, which is quite specific.

\subsection{Richardson equations}

It was shown by Richardson that the Hamiltonian, defined in Eqs.
(\ref{BCSpot}) and (\ref{H0}), is exactly solvable. The energy of
$N$ pairs is given by the
sum of $N$ energy-like complex quantities $R_{j}$ ($j=1$,..., $N$)%
\begin{equation}
E_{N}=\sum_{j=1}^{N}R_{j}.  \label{sum}
\end{equation}%
These quantities satisfy the system of $N$ coupled nonlinear
algebraic
equations, called Richardson equations. The equation for $R_{j}$ reads as%
\begin{equation}
1=\sum_{\mathbf{k}}\frac{V}{2\varepsilon
_{\mathbf{k}}-R_{j}}+\sum_{l,l\neq j}\frac{2V}{R_{j}-R_{l}},
\label{Richardson}
\end{equation}
where the summation in the first term of the RHS of the above
equation is performed for $\varepsilon _{\mathbf{k}}$ located in
the energy interval, where the potential acts. Note that the
number of pairs enters to the formalism
through the number of equations that is rather unusual. The case $N=1$ at $%
\Lambda \rightarrow \infty $, corresponds to the one-pair problem
solved by Cooper. The fully analytical resolution of Richardson
equations, in general case, stays an open problem.

\subsection{Electrostatic analogy}

Let us consider the function $E_{class}(\left\{ R_{j}\right\} )$,
given by
\begin{equation}
E_{class}(\left\{ R_{j}\right\} )=2\left(
\sum_{j}ReR_{j}+V\sum_{j,\mathbf{k}}\ln \left\vert 2\varepsilon
_{\mathbf{k}}-R_{j}\right\vert -2V\sum_{j,l,j<l}\ln \left\vert
R_{l}-R_{j}\right\vert \right) .  \label{EnergyCharges}
\end{equation}%
This function can be rewritten as%
\begin{equation}
E_{class}(\left\{ R_{j}\right\} )=W(\left\{ R_{j}\right\}
)+W(\left\{ R_{j}^{\ast }\right\} ),  \label{WW*}
\end{equation}%
where%
\begin{equation}
W(\left\{ R_{j}\right\} )=\sum_{j}R_{j}+V\sum_{j,\mathbf{k}}\ln
\left( 2\varepsilon _{\mathbf{k}}-R_{j}\right)
-2V\sum_{j,l,j<l}\ln \left( R_{l}-R_{j}\right) .  \label{W}
\end{equation}%
Richardson equations can be formally written\cite{Rich3} as
stationary conditions for $W(\left\{ R_{j}\right\} )$: $\partial
W(\left\{ R_{j}\right\} )/\partial R_{j}=0$. It is easy to see
that $E_{class}(\left\{ R_{j}\right\} )$ represents an energy of
$N$ free classical particles with electrical charges $2\sqrt{V}$
located on the plane with coordinates given by (Re $R_{j} $, Im
$R_{j}$). These particles are subjected into an external uniform
force directed along the axis of abscissa with the strength $-2$.
They are attracted to $N_{\Omega }$ fixed particles each having a
charge $-\sqrt{V}$ and located at $\varepsilon _{\mathbf{k}}$'s.
Free charges repeal each other. Richardson equations are
equivalent to the equilibrium condition for the system of $N$ free
charges.

\subsection{"Probabilistic" approach}

A key idea of the approach we here suggest is to switch from the
"energy" of the system of charges to the "probability" $S(\left\{
R_{j}\right\} )$ for this system to be in a certain state at
effective "temperature" $T_{eff}$ given by the simple condition
\begin{equation}
k_{B}T_{eff}\equiv V,  \label{Tfict}
\end{equation}%
where%
\begin{equation}
S(\left\{ R_{j}\right\} )=\exp \left( -\frac{W(\left\{ R_{j}\right\} )}{V}%
\right).  \label{S}
\end{equation}%
Taking into account Eq. (\ref{W}) for $W(\left\{ R_{j}\right\} )$,
we obtain
for $S(\left\{ R_{j}\right\} )$\ a nicely compact expression%
\begin{equation}
S(\left\{ R_{j}\right\} )=\frac{\prod_{j,l,j<l}(R_{l}-R_{j})^{2}}{%
\prod_{j=1}^{N}\prod_{\mathbf{k}}(2\varepsilon
_{\mathbf{k}}-R_{j})}\exp \left(
-\frac{\sum_{j=1}^{N}R_{j}}{V}\right),  \label{Sproduct}
\end{equation}
which has obvious similarities with the square of Laughlin wave
function at filling 1, due to the factor
$\prod_{j,l,j<l}(R_{l}-R_{j})^{2}$.

In principle, it could seem more reasonable to use
$E_{class}(\left\{ R_{j}\right\} )$ instead of $W(\left\{
R_{j}\right\} )$ in the definition of "probability". Indeed, the
corresponding function $\exp \left( -E_{class}(\left\{
R_{j}\right\} )/V\right)$ is real-valued and positive, so that by
its properties it is closer to a usual distribution function, as
compared to $S(\left\{ R_{j}\right\} )$. However, such a function
is not meromorphic, so that Cauchy theorem is not applicable;
therefore, it is not so useful for the reasons, which will be
clarified below.

The important fact is that the effective "temperature" $T_{eff}$
goes to zero in the thermodynamical limit as $k_{B}T_{eff}\sim
\Lambda ^{-1}$. This makes $W(\left\{ R_{j}\right\} )/V$ extremely
large by its absolute value, while the landscape of $S(\left\{
R_{j}\right\} )$ is very sharp in the vicinity of stationary
points of $W(\left\{ R_{j}\right\} )$. Hence, it looks attractive
to try extracting an information about stationary points of
$W(\left\{ R_{j}\right\} )$\ by using integration techniques. We
briefly illustrate this idea. Let $g(x)$ be a function of the
variable $x$, which has a sharp maximum at $x=x_{0}$. It is
possible to find $x_{0}$ approximately without solving equation
$g^{\prime }(x_{0})=0$ explicitly, but through the integration.
Namely, we consider a ratio of integrals $\int x\exp (g(x))dx/\int
\exp (g(x))dx$ with $x_{0}$ located "deeply enough" inside the
integration interval. The dominant contribution to both integrals
is provided by a neighborhood of $x_{0}$, so that we expect that
their ratio
is close to $x_{0}$. In particular, if a second and higher-order derivatives of $%
g(x)$ at $x_{0}$ are proportional to some large parameter, for example to $%
\Lambda $, then it is easy to show, by performing Taylor expansion
of $g(x)$ around $x_{0}$, that the error in the determination of
$x_{0}$ through the above ratio is of the order of $\Lambda
^{-1}$.

However, it is easily seen that stationary points do not
necessarily correspond to minima of "energy", but they rather give
its saddle points, so that equilibrium positions of free charges
are not stable. The easiest way of seeing it is to consider a
one-pair problem, for which equilibrium $R$ is located on the real
axis, as indicated in Fig. 1. The energy of the system
$E_{class}(R)$ (as well as the real part of $W(R)$) increases, if
we start to move $R$ out of equilibrium from the real axis in a
perpendicular direction, but it decreases upon motion along the
real axis in both directions (see Section IV for more details).
Hence, to apply a saddle-point method, we should use an
integration path passing through the stationary point, as shown in
Fig. 1 by line 1. This procedure could be seen as useless from the
viewpoint of determination of a saddle point position, since we
already need to know it to apply this technique. However, this is
not true, because we use $W(\left\{ R_{j}\right\} )$ instead of
$E_{class}(\left\{ R_{j}\right\} )$ in the definition of the
"probability" $S$. Function $S$ is meromorphic, which means that
the result of integration is independent on an integration path
for all paths which can be continuously transformed to each other
without crossing any pole of $S$. Hence, we can use a variety of
paths, which start at $-i\infty$ and end at $+i\infty$, where $S$
vanishes. \textit{By using these "nonlocal" properties of $S$, we
can reconstruct an information on the unknown location of a saddle
point in the complex plane}. Therefore, we are led to consider
the ratio of integrals defined as%
\begin{equation}
E\equiv \frac{\int RS(R) dR}{\int S(R) dR}%
,  \label{Eint}
\end{equation}%
where integration is performed in a complex plane from $%
-i\infty $ to $+i\infty $. From one point of view, by passing
integration path through the saddle point, we must recover with
accuracy $\sim 1/\rho$ the position of the saddle point. From
another point of view, we may use any integration path, provided
it avoids all poles in the same way as line 1 of Fig. 1. Taking
into account Eq. (\ref{Sproduct}) for $S(\left\{ R_{j}\right\} )$,
we may
rewrite Eq. (\ref{Eint}) in an equivalent form as%
\begin{equation}
E=-\frac{\partial }{\partial \left( \frac{1}{V}\right) }\ln Z,
\label{logderiv}
\end{equation}%
where
\begin{equation}
Z\equiv \int S(R) dR. \label{partit}
\end{equation}%

\begin{figure}[tbp]
\begin{center}
\includegraphics[width=0.5\textwidth]{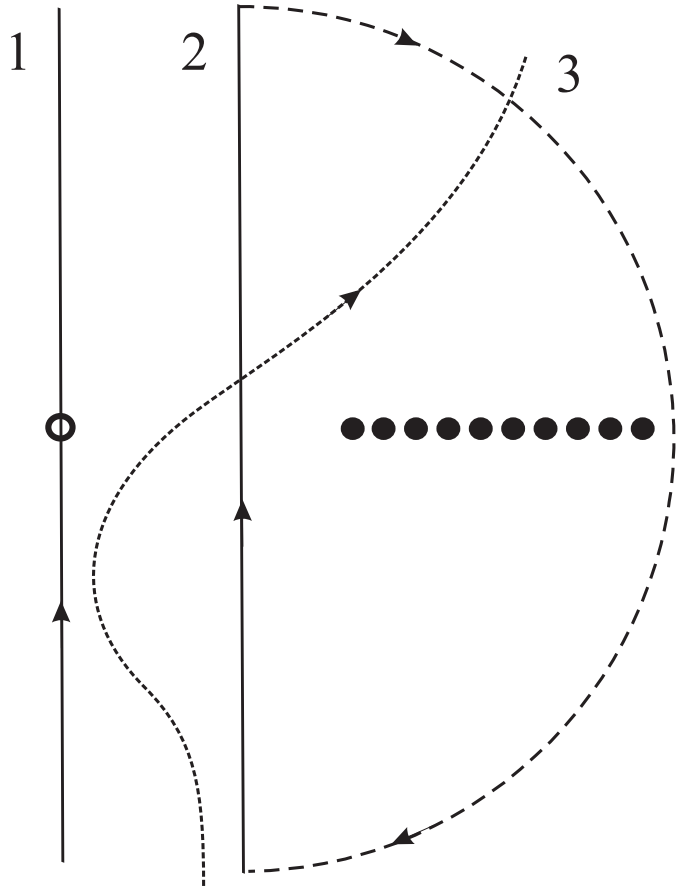}
\end{center}
\caption{The schematic plot of complex values of $R$ relevant for
a single-pair problem. Filled circles show locations of
one-electronic levels. Open circle indicates an equilibrium
position of $R$ for the ground state. Line 1 shows an integration
path through this point in the direction of the steepest descent
of $-$Re $W$. Lines 2 and 3 represent examples of integration
paths, for which the results of integration will be the same as
for the line 1. Dashed line depicts the auxiliary semicircle
introduced to use a residue theorem.} \label{Fig1}
\end{figure}

For a single-pair problem, $S$ reads as
\begin{equation}
S(R)=\frac{\exp (-R/V)}{\prod_{n=0}^{N_{\Omega }}(2\varepsilon _{F_{0}}+%
\frac{2n}{\rho }-R)}.  \label{Ssingle}
\end{equation}%
We then perform a partial-fraction decomposition of
$1/\prod_{n=0}^{N_{\Omega
}}(2\varepsilon _{F_{0}}-R+2n/\rho )$ by using standard rules and rewrite $S$ as%
\begin{equation}
S(R)=C\exp (-2\varepsilon _{F_{0}}/V)\sum_{n=0}^{N_{\Omega
}}(-1)^{n}\binom{N_{\Omega }}{n}\frac{\exp (-R/V)}{2\varepsilon
_{F_{0}}-R+\frac{2n}{\rho }},  \label{Ssb}
\end{equation}%
where $C$ is an irrelevant $V$-independent constant ($C=\left(
\rho /2\right) ^{N_{\Omega }}(N_{\Omega }!)^{-1}$), which will be
dropped below, as well as similar irrelevant factors, while
\begin{equation}
\binom{N_{\Omega }}{n}=\frac{N_{\Omega }!}{n!(N_{\Omega }-n)!}
\end{equation}%
is a binomial coefficient. We then substitute Eq. (\ref{Ssb}) to Eq. (\ref%
{partit}) and perform an integration. This can be done by various
methods. For instance, we may perform integration in a
straightforward way, for the integration path along a straight
line, like line 2 in Fig. 1. We then see that, of course, the
result of the integration depends only on how many poles of
function $S$, as given by Eq. (\ref{Ssb}) (corresponding to the
positions of one-electronic levels $2\varepsilon _{\mathbf{k}}$ in
the complex plane), is at the right (or left) side of the
integration path and it stays the same as long as the number of
poles at the right side is the same. In particular, the result is
identical for the lines 1 and 2. Alternatively, we can use a
residue theorem. For the integration contour, we choose a path,
which consists of the line going from $x-iy$ to $x+iy$ (for
instance, line 2 in Fig. 1) and then clockwises along a
semicircle, as again shown in Fig. 1. We then consider a limit
$y\rightarrow \infty $, which ensures that $S$ is infinitely small
along the auxiliary semicircle, as follows from Eq.
(\ref{Ssingle}). The role of this semicircle is purely technical:
it is needed to apply a residue theorem. The result of the
integration depends only on how many poles of $S(R)$ are enclosed
by the contour. By enclosing the whole set of poles, we arrive to
$Z$ corresponding to the ground state of the initial
quantum-mechanical problem, for which an equilibrium $R$ is
located to the left of the whole set of the one-electronic
energies on a complex plane. Enclosing less number of poles leads
to $Z$ for excited states, for which an equilibrium $R$ is known
to be positioned between two one-electronic levels on real axis
(quasi-continuum spectrum). Excited states are beyond the scope of
the present paper and will be addressed elsewhere. Note that it is
not necessary at all to consider paths consisting of straight
lines (see, e.g., curve 3 in Fig. 1).

After integration, we readily
obtain a very simple expression for the ground-state $Z$%
\begin{equation}
Z=\exp (-2\varepsilon _{F_{0}}/V)\sum_{n=0}^{N_{\Omega }}(-1)^{n}\sigma ^{n}%
\binom{N_{\Omega }}{n}=\exp (-2\varepsilon _{F_{0}}/V)\left(
1-\sigma \right) ^{N_{\Omega }}.  \label{binom}
\end{equation}%
By finding a logarithmic derivative of $Z$ with respect to $1/V$,
we recover
a well-known expression of the single-pair energy%
\begin{equation}
E_{1}=2\varepsilon _{F_{0}}-2\Omega \frac{\sigma }{1-\sigma },
\label{pairener}
\end{equation}
which also coincides with Eq. (1) for $N=1$.

This method yields the single-pair energy with error of the order
of $1/\rho $. The same however applies to the traditional method
to derive Eq. (\ref{pairener}), which relies on the replacement of
the sum by the integral (see Section IV).

The suggested scheme can be applied to the case of many pairs. Let us
restrict ourselves for the moment to values of $N$ such that $N \leq
N_{\Omega }/2$ for which the degree of the polynomial
$\prod_{l,j,j<l}(R_{l}-R_{j})^{2}$ in each $R_{j}$ is smaller than
that of the polynomial
$\prod_{j=1}^{N}\prod_{\mathbf{k}}(2\varepsilon
_{\mathbf{k}}-R_{j})$. It then follows from Eq. (\ref{Sproduct}) that $%
S(\left\{ R_{j}\right\} )$, $S(\left\{ R_{j}\right\}
)\sum_{j=1}^{N}R_{j}\rightarrow 0$ at Im $R_{n}\rightarrow \pm
\infty $ for any $R_{n}$ from the set $\left\{ R_{j}\right\} $.

We consider a system of free charges in the equilibrium and then
start to move this system \textit{as a whole} out of equilibrium,
in such a way that mutual distances between free charges stay the
same, while the position of the center of mass $\bar{R}$ of this
system changes. Now, we concentrate on the change $\delta W$ of
$W(\left\{ R_{j}\right\} )$ upon this motion. By expanding
$W(\left\{ R_{j}\right\} )$ in powers of the deviation $\delta
\bar{R}=\bar{R}-\bar{R}^{(0)}$ of $\bar{R}$ from the equilibrium,
we easily arrive to the equation
\begin{equation}
\delta W=V\sum_{n=2}^{\infty}\kappa_{n}(\delta \bar{R})^{n},
\label{centerofmass}
\end{equation}
where
\begin{equation}
\kappa_{n}=-\frac{1}{n} \sum_{j, \mathbf{k} } \frac{1}{(
2\varepsilon _{\mathbf{k}}-R_{j}^{(0)})^{n}}, \label{kappa}
\end{equation}
where $R_{j}^{(0)}$ are positions of free charges in equilibrium.
Next, we assume that in the ground state all $R_{j}^{(0)}$ are
located far enough from the line of fixed charges, so that
distances between each free charge and this line are much larger
than $1/\rho$. This assumption is rather natural for the
thermodynamical limit. Besides, the method for the solution of
Richardson equations developed earlier in Ref. [\cite%
{Rich3}] is also based on the same assumption, since it utilizes a
continuous approximation for the positions of free charges.
However, in contrast to [\cite {Rich3}], we do not introduce any
additional requirements on the \textit{shape} of a distribution of
$R_{j}^{(0)}$ on the complex plane.

In the limit $N \rightarrow \infty$, it follows directly from Eq.
(\ref{kappa}) that $\kappa_{n}$ depends on system volume as
\begin{equation}
\kappa_{n} \sim \rho N \Omega^{1-n}. \label{kappascale}
\end{equation}
Moreover, $\kappa_{2}$ is purely real due to the mirror symmetry
of the system of charges with respect to the real axis. Eq.
(\ref{kappascale}) implies that the absolute value of $S$ is
extremely strongly peaked near the equilibrium. Namely, if we go
in the direction of the steepest descent of $-$Re $W$, then it is
peaked within the interval of the width $\sim 1/\rho$. If we now
consider a ratio of integrals similar to the one in Eq.
(\ref{Eint}), where the integration is performed for the position
of the center of mass as well as for internal degrees of freedom
of the system of charges, then we can find the equilibrium
location of the center of mass, while the error will be in powers
in $1/N$, as Eq. (\ref{kappascale}) suggests, so that it is
negligible in the thermodynamical limit. We may formally extend
the integration path for $\bar{R}$ from $-i\infty$ to $+i\infty$,
where $S$ is zero. At the same time, the integrals entering the
ratio are equivalent to the multidimensional integrals with
respect to all $R$'s. By using the Cauchy theorem, we may deform
the integration path for each $R$, limits of integration being
$-i\infty$ and $+i\infty$. Thus, we are led to consider the ratio
of integrals defined as in Eq. (\ref{Eint}), where an integration
is performed for each $R_{j}$. The ratio $E$ will allow us to
reconstruct an information about the sum of all $R$'s in an
equilibrium, while this is the quantity of interest, since it is
equal to the energy of the initial quantum-mechanical system. It
is again convenient to introduce $Z$ given by Eq. (\ref{partit}),
where an
integration is now performed for all variables $R_{j}$ from $%
-i\infty $ to $+i\infty $. The sum of equilibrium $R$'s is also
given by Eq. (\ref{logderiv}). Eqs. (\ref{logderiv}) and
(\ref{partit}) lead us to associate $Z$ with the "partition
function" for our classical system of charges, although such an
analogy is certainly not complete. Nevertheless, we think that
this term is far from being meaningless, in the view of Eq.
(\ref{logderiv}). A nontrivial feature is that the energy of the
initial quantum mechanical system is determined by the
logarithmical derivative of this classical "partition function".

Note that our approach has some analogies with thermodynamics.
Indeed, for the system of many particles, it is often hopeless to
resolve equations for equilibrium positions of each particle.
However, such a detailed information is not necessary to
understand many global properties of the system, so that a
thermodynamical description becomes highly efficient. The major
difference with the usual thermodynamics is in the fact that we
are dealing with the unstable equilibrium and therefore the
"partition function" is obtained by integrating only over half of
the degrees of freedom and by using a complex-valued
"probability". We also would like to note that the method we here
suggest turns out to have similarities with the large-$N$
expansion for the two-dimensional Dyson gas proposed recently in
Ref. [\cite{Zabrodin}]. Perhaps, our approach may be also related
to inverse problems in mathematics. For example, there exists the
Radon transform, which allows one to reconstruct an unknown
function by using integrals of this function along various paths
[\cite{Radon}]. This method is widely used in the tomography.

One-dimensional integrals having denominator of the integrand of the form $%
\prod_{k=0}^{N_{\Omega}}(r-k)$, while integration for $r$ is performed from $%
-i\infty $ to $+i\infty $ are called N\"{o}rlund-Rice integrals.
Here we deal with the multidimensional coupled integrals of
N\"{o}rlund-Rice type. At the same time, factor
$\prod_{l,j,j<l}(R_{l}-R_{j})^{2}$ in the integrand makes them
similar to Selberg integrals. Note that it is actually known that
N\"{o}rlund-Rice integrals can be transformed to binomial sums and
vise versa\cite{Sedgewick}.

In the next Section, we apply a technique based on binomial sums for
finding the ground state energy for arbitrary $N$. However, before
doing this, let us address one more point.

We have imposed a restriction $N\leq N_{\Omega }/2$. Configurations with $%
N\geq N_{\Omega }/2$ can be handled by using a concept of holes,
i.e., empty states in the potential layer. In the Appendix A, we
show that the initial Hamiltonian is characterized by a duality
between electrons and holes, which
allows us to map states with $N<N_{\Omega }/2$ to the states with $%
N>N_{\Omega }/2$.

\section{Ground state energy through the binomial sum}

In this Section, we consider the ground state energy for $N$
pairs, $N\leq
N_{\Omega }/2$. It is seen from Eq. (\ref{Sproduct}) that the "probability" $%
S$ can be rewritten using a partial-fraction decomposition, in a
manner,
similar to the one used for the one-pair problem, as%
\begin{equation}
S(\left\{ R_{j}=2\varepsilon _{F_{0}}-r_{j}\right\} )=\exp \left( \frac{%
-2N\varepsilon _{F_{0}}}{V}\right)
\left[\prod_{l,j,l>j}(r_{l}-r_{j})^{2}\right]
\sum_{n_{1},n_{2}...,n_{N}=0}^{N_{\Omega }}
 \left[
\prod_{j=1}^{N}(-1)^{n_{j}}\binom{N_{\Omega }}{n_{j}}\frac{\exp
(r_{j}/V)}{r_{j}+\frac{2n_{j}}{\rho }}\right]. \label{SN}
\end{equation}%
It can be of interest to note that the sum given by Eq. (\ref{SN})
may be rewritten in terms of the forward difference operators, as
usual for binomial sums\cite{umbral,Sedgewick}. This means that
Richardson equations can be represented in a similar way too.

After substitution of Eq. (\ref{SN}) to Eq. (\ref{partit}) and
integrating in such a way that the integration path for each $R$
avoids the whole set of poles, as for the one-pair problem,
we obtain the binomial sum, given by%
\begin{equation}
Z=\exp (-2N\varepsilon _{F_{0}}/V)z,  \label{zsmall}
\end{equation}%
where
\begin{equation}
z=\sum_{n_{1},n_{2}...,n_{N}=0}^{N_{\Omega }}\left[
\prod_{j=1}^{N}(-1)^{n_{j}}\sigma ^{n_{j}}\binom{N_{\Omega
}}{n_{j}}\right] \prod_{l,j,l>j}(n_{l}-n_{j})^{2}.  \label{binom1}
\end{equation}%
This sum again corresponds to the ground state of the
quantum-mechanical system, for which none of equilibrium $R$'s is
located between two one-electronic levels in the real axis.

Without the last factor in the RHS of Eq. (\ref{binom1}), this sum
reduces to the product of trivial binomial sums for each $n_{j}$,
as the one of Eq. (\ref{binom}). In order to tackle the sum with
the coupling factor, we first provide some identities, which will
be very useful in the derivation presented below.

It is convenient to introduce Pochhammer symbol (or falling
factorial) given by
\begin{equation}
(n)_{q}=n(n-1)...(n-q+1),  \label{pochh}
\end{equation}%
while $(n)_{0}\equiv 1$. It is then easy to obtain the following
identity
\begin{equation}
z_{q,p}\equiv \sum_{n=0}^{N_{\Omega }}(-1)^{n}\sigma ^{n}\binom{N_{\Omega }}{%
n}(n)_{q}(N_{\Omega }-n)_{p}=(-1)^{q}\sigma ^{q}(1-\sigma )^{N_{\Omega }-q-p}%
\frac{N_{\Omega }!}{(N_{\Omega }-q-p)!},  \label{eat}
\end{equation}%
where $q+p\leq N_{\Omega }$. If we now consider a product of sums for each $%
n_{j}$, every sum being similar to the one of Eq. (\ref{eat}), we
get
\begin{eqnarray}
z_{q_{1},p_{1}}...z_{q_{N},p_{N}}
&=&\sum_{n_{1},n_{2}...,n_{N}=0}^{N_{\Omega
}}\prod_{j=1}^{N}(-1)^{n_{j}}\sigma ^{n_{j}}\binom{N_{\Omega }}{n_{j}}%
(n_{j})_{q_{j}}(N_{\Omega }-n_{j})_{p_{j}}  \notag \\
&=&\sigma ^{\sum_{j=1}^{N}q_{j}}(1-\sigma )^{NN_{\Omega
}-\sum_{j=1}^{N}(q_{j}+p_{j})}(-1)^{\sum_{j=1}^{N}q_{j}}\prod_{j=1}^{N}\frac{%
N_{\Omega }!}{(N_{\Omega }-q_{j}-p_{j})!}.  \label{eating}
\end{eqnarray}%
Note that (i) only first two factors in the last line of Eq.
(\ref{eating}) depend on the interaction constant $V$ (through
$\sigma $); (ii) the
dependence of both these factors on the two \textit{sets} of numbers $%
\left\{ q_{j}\right\} $ and $\left\{ p_{j}\right\} $ is only through $%
\sum_{j=1}^{N}q_{j}$ and $\sum_{j=1}^{N}p_{j}$, which are just
degrees of polynomials $\prod_{j=1}^{N}(n_{j})_{q_{j}}$ and
$\prod_{j=1}^{N}(N_{\Omega }-n_{j})_{p_{j}}$, respectively. These
two observations turn out to be of a crucial importance.

Our approach is to transform the initial coupled sum, as given by Eq. (\ref%
{binom1}), into a linear combination of uncoupled sums, similar to
the one given by Eq. (\ref{eating}). To make such an uncoupling,
we note that, as known, $\prod_{l,j,l>j}(n_{l}-n_{j})$ can be
rewritten as the determinant of the Vandermonde matrix
\begin{equation}
\prod_{l,j,l>j}(n_{l}-n_{j})\equiv V(\left\{ n_{j}\right\} )=\det
\begin{bmatrix}
1 & 1 & 1 & \ldots & 1 \\
n_{1} & n_{2} & n_{3} & \ldots & n_{N} \\
n_{1}^{2} & n_{2}^{2} & n_{3}^{2} & \ldots & n_{N}^{2} \\
\ldots & \ldots & \ldots & \ldots & \ldots \\
n_{1}^{N-1} & n_{2}^{N-1} & n_{3}^{N-1} & \ldots & n_{N}^{N-1}%
\end{bmatrix}%
.  \label{vander}
\end{equation}

Next step is to note that $\prod_{l,j,l>j}(n_{l}-n_{j})$ can be
\textit{also}
written as $(-1)^{N(N-1)/2}V(\left\{ N_{\Omega }-n_{j}\right\} )$, since $%
n_{l}-n_{j}=-\left[ (N_{\Omega }-n_{l})-(N_{\Omega }-n_{j})\right]
$. Hence, we obtain the identity
\begin{equation}
\prod_{l,j,l>j}(n_{l}-n_{j})^{2}=(-1)^{N(N-1)/2}V(\left\{
{n_{j}}\right\} )V(\left\{ N_{\Omega }-n_{j}\right\} ).
\label{split}
\end{equation}

Now we make use of the well-known rule that the determinant of the
matrix does not change if we add a multiple of one row to another
row. It is then easy to see that $V(\left\{ {n_{j}}\right\} )$ can
be rewritten in a "falling factorial" form as
\begin{equation}
V(\left\{ {n_{j}}\right\} )=\det
\begin{bmatrix}
(n_{1})_{0} & (n_{2})_{0} & (n_{3})_{0} & \ldots & (n_{N})_{0} \\
(n_{1})_{1} & (n_{2})_{1} & (n_{3})_{1} & \ldots & (n_{N})_{1} \\
(n_{1})_{2} & (n_{2})_{2} & (n_{3})_{2} & \ldots & (n_{N})_{2} \\
\ldots & \ldots & \ldots & \ldots & \ldots \\
(n_{1})_{N-1} & (n_{2})_{N-1} & (n_{3})_{N-1} & \ldots & (n_{N})_{N-1}%
\end{bmatrix}%
,  \label{vander1}
\end{equation}%
while $V(\left\{ N_{\Omega }-n_{j}\right\} )$ can be represented
in a similar form with $n_{j}$ changed into $N_{\Omega }-n_{j}$.

It is obvious that, using Eq. (\ref{vander1}), $V(\left\{
{n_{j}}\right\} )$
can be written as a linear combination of polynomials each having a form $%
\prod_{j=1}^{N}(n_{j})_{q_{j}}$. The crucial point is that for
each term in this sum, $\sum_{j=1}^{N}q_{j}$ \textit{is the same}:
it is just equal to the degree of the polynomial $V(\left\{
{n_{j}}\right\} )$. This degree is
equal to the sum of degrees of polynomials from each row that is to $%
0+1+...+(N-1)=N(N-1)/2$. The same applies to $V(\left\{ N_{\Omega
}-n_{j}\right\} )$, which is represented as a linear combination
of
polynomials of the form $\prod_{j=1}^{N}(N_{\Omega }-n_{j})_{p_{j}}$ with $%
\sum_{j=1}^{N}p_{j}=N(N-1)/2$. Now we see that Eq. (\ref{vander1})
together with the similar equation for $V(\left\{ N_{\Omega
}-n_{j}\right\} )$ allows us to rewrite
$\prod_{l,j,l>j}(n_{l}-n_{j})^{2}$ as a linear combination of
polynomials of the form $\left(
\prod_{j=1}^{N}(n_{j})_{q_{j}}\right) \left(
\prod_{j=1}^{N}(N_{\Omega }-n_{j})_{p_{j}}\right) $ with \textit{the same} $%
\sum_{j=1}^{N}q_{j}$ and $\sum_{j=1}^{N}p_{j}$ for each
polynomial. At this stage, we immediately apply Eq. (\ref{eating})
and obtain
\begin{equation}
z=\sigma ^{\frac{N(N-1)}{2}}(1-\sigma )^{NN_{\Omega
}-N(N-1)}A(N,N_{\Omega }),  \label{Zfinal}
\end{equation}%
where $A(N,N_{\Omega })$ is some function of $N$ and $N_{\Omega
}$, which is irrelevant for the determination of the
quantum-mechanical energy, since the later is given by the
logarithmic derivative of $Z$ with respect to $1/V$. Finally, by
finding this logarithmic derivative and by taking into account Eq.
(\ref{zsmall}), we easily arrive to Eq. (1) for the ground state
energy.

Eq. (\ref{Zfinal}) can be derived using more formal way of
writing. We first note that $V(\left\{ {n_{j}}\right\} )$ can be
expressed in the following form
\begin{eqnarray}
V(\left\{ {n_{j}}\right\} )\equiv\det(V_{q,p})=\det[(n_{p})_{q-1}]  \notag \\
=\sum_{j_{1},j_{2},...,j_{N}=1}^{N}
\varepsilon_{j_{1}j_{2}...j_{N}}(n_{j_{1}})_{0}(n_{j_{2}})_{1}...
(n_{j_{N}})_{N-1},  \label{LeviCivita}
\end{eqnarray}
where $\varepsilon_{j_{1}j_{2}...j_{N}}$ is the Levi-Civita
symbol. Similarly, we can write
\begin{eqnarray}
V(\left\{ {n_{j}}\right\} )V(\left\{ N_{\Omega}-{n_{j}}\right\} )
=\sum_{j_{1},j_{2},...,j_{N}=1}^{N}
\varepsilon_{j_{1}j_{2}...j_{N}}
\sum_{j_{1}^{\prime},j_{2}^{\prime},...,j_{N}^{\prime}=1}^{N}
\varepsilon_{j_{1}^{\prime}j_{2}^{\prime}...j_{N}^{\prime}}  \notag \\
(n_{j_{1}})_{0}(n_{j_{2}})_{1}...(n_{j_{N}})_{N-1}
(N_{\Omega}-n_{j_{1}^{\prime}})_{0}(N_{\Omega}-n_{j_{2}^{\prime}})_{1}...
(N_{\Omega}-n_{j_{N}^{\prime}})_{N-1}.  \label{2LeviCivita}
\end{eqnarray}
By definition, the Levi-Civita symbol is nonzero only for the set
of its indices all different. This means that, for nonzero terms
in the double sum in the
RHS of Eq. (\ref{2LeviCivita}), there should be no repetitions in the set $%
j_{1},j_{2},...,j_{N}$ and also in another set $j_{1}^{\prime},j_{2}^{%
\prime},...,j_{N}^{\prime}$. We now use Eq. (\ref{eating}) and see
that each nonzero term of the double sum gives the same dependence
on $V$, when substituted into Eq. (\ref{zsmall}). Thus, we again
arrive to Eq. (\ref{Zfinal}).

In order to reach some qualitative understanding of the result, we
have obtained, let us come back to the expression of
"probability", as given by
Eq. (\ref{Sproduct}). We rewrite it as%
\begin{eqnarray}
S(\left\{ R_{j}=2\varepsilon _{F_{0}}-r_{j}\right\} ) &=&\exp
(-2N\varepsilon
_{F_{0}}/V)\frac{1}{\prod_{j=1}^{N}\prod_{n=0}^{\rho \Omega
}(r_{j}+\frac{2n}{\rho })}\exp \left(
\frac{\sum_{j=1}^{N}r_{j}}{V}\right)
\notag \\
&&\det
\begin{bmatrix}
1 & 1 & \ldots & 1 \\
\prod_{n=0}^{0}(r_{1}+\frac{2n}{\rho }) & \prod_{n=0}^{0}(r_{2}+\frac{2n}{%
\rho }) & \ldots & \prod_{n=0}^{0}(r_{N}+\frac{2n}{\rho }) \\
\ldots & \ldots & \ldots & \ldots \\
\prod_{n=0}^{N-2}(r_{1}+\frac{2n}{\rho }) & \prod_{n=0}^{N-2}(r_{2}+\frac{2n%
}{\rho }) & \ldots & \prod_{n=0}^{N-2}(r_{N}+\frac{2n}{\rho })%
\end{bmatrix}
\notag \\
&&\det
\begin{bmatrix}
1 & 1 & \ldots & 1 \\
\prod_{n=0}^{0}(r_{1}+2\Omega -\frac{2n}{\rho }) & \prod_{n=0}^{0}(r_{2}+2%
\Omega -\frac{2n}{\rho }) & \ldots & \prod_{n=0}^{0}(r_{N}+2\Omega -\frac{2n%
}{\rho }) \\
\ldots & \ldots & \ldots & \ldots \\
\prod_{n=0}^{N-2}(r_{1}+2\Omega -\frac{2n}{\rho }) &
\prod_{n=0}^{N-2}(r_{2}+2\Omega -\frac{2n}{\rho }) & \ldots &
\prod_{n=0}^{N-2}(r_{N}+2\Omega -\frac{2n}{\rho })%
\end{bmatrix}%
.  \label{factoriz}
\end{eqnarray}

It is seen from the above equation that the "probability"
$S(\left\{ R_{j}\right\} )$ can be represented in a factorized
form as a sum of
products of $N$ "probabilities", each of them being the one for a \textit{%
single pair}. Each single pair is however placed into its own
environment with a band of one-electronic states removed from the
bottom of the potential layer and another band of states removed
from the top of the potential layer, as a result of the Pauli
exclusion principle. The energy of a single pair in the state with
$n$ levels removed from the bottom of the potential layer
and $m$ states removed from the top of the layer is obtained from Eq. (\ref%
{pairener}) by substitution $\varepsilon _{F_{0}}\rightarrow
\varepsilon _{F_{0}}+n/\rho $, $\Omega\rightarrow \Omega
-(n+m)/\rho $. Hence, it follows from Eq. (\ref{factoriz}) that
the \textit{sum} of single-pair energies is the same for each term
of the factorized "probability", although sets of single-pair
energies for different terms are not identical.

We therefore can think that the original system of $N$ pairs,
feeling each other through the Pauli exclusion principle, is a
superposition of "states" of $N$ single pairs, but each pair is
placed into its own environment with bands of the states deleted
both from the bottom and from the top of the potential layer.
Moreover, the sum of energies of these single pairs for each
"state" of the superposition is the same, which is, probably, a
consequence of the constant density of states. This understanding
provides an additional nontrivial link between the one-pair
problem, solved by Cooper, and many-pair BCS theory.

Note that we see here a quite close
analogy with the well-known Hubbard-Stratonovich transformation [\cite%
{Strat,Hub}], which enables one to represent the partition
function for the system of interacting particles through the
partition function for the system of noninteracting (single)
particles, but in the fluctuating field. We also would like to
mention that some of the "probabilities" appearing in Eq.
(\ref{factoriz}) can be negative; one should not be confused by
this fact, since negative "probabilities" are known to appear in
problems involving fermions. In particular, they lead to the
well-known negative sign problem arising when trying to apply
Monte Carlo methods to compute partition functions. This problem
becomes severe in the thermodynamical limit. Actually, we here see
some reminiscence of the same problem. Namely, if we try to
integrate factorized $S$ through paths crossing individual saddle
points of each single pair, we immediately see that the absolute
value of resulting $Z$ is going to be much smaller than the result
of the integration for each term due to the "determinantal"
structure of Eq. (\ref {factoriz}). The easiest way to see it is
to consider the two-pair problem. The situation is getting more
and more difficult with the increase of $N$. Fortunately, by
applying the method based on binomial sums, we can circumvent all
these difficulties and to evaluate $Z$ exactly.

\section{Ground state energy through the N\"{o}rlund-Rice integral}

In this Section, we present another method, which allows us to
find the ground state energy as an expansion in powers of pairs
density. In practise, this method is much less powerful compared
to the one described in the previous Section. We here calculate
just few first terms of the expansion, since calculations become
more and more heavy with increasing number of terms. One of the
main aims of this Section is actually to establish a link
between our approach and the dilute-limit approach of Refs. [\cite{We,EPJB}%
]. The method presented in this Section turns out to be closely
related to the method of Refs. [\cite{We,EPJB}].

The general idea is not to transform the initial multidimensional
integral of N\"{o}rlund-Rice type into a binomial sum, but to
tackle it by using a saddle-point method. For $z$, we have
\begin{equation}
z=\int dr\frac{\prod_{l,j,j<l}(r_{l}-r_{j})^{2}}{\prod_{j=1}^{N}%
\prod_{m_{j}=0}^{N_{\Omega }}(r_{j}+\frac{2m_{j}}{\rho })}\exp \left( \frac{%
\sum_{j=1}^{N}r_{j}}{V}\right) ,  \label{ZNorlund}
\end{equation}%
where%
\begin{equation*}
\int dr=\int dr_{1}\ldots \int dr_{N}.
\end{equation*}%
Quantum-mechanical energy $E$ is expressed through the logarithmic
derivative of $z$ with respect to $1/V$ as
\begin{equation}
E\equiv 2N\varepsilon _{F_{0}}-\frac{\partial ln(z)}{\partial (1/V)}%
=2N\varepsilon _{F_{0}}-\frac{1}{z}\int dr\left(
\sum_{j=1}^{N}r_{j}\right)
\frac{\prod_{l,j,j<l}(r_{l}-r_{j})^{2}}{\prod_{j=1}^{N}\prod_{m_{j}=0}^{N_{%
\Omega }}(r_{j}+\frac{2m_{j}}{\rho })}\exp \left( \frac{\sum_{j=1}^{N}r_{j}}{V}%
\right)  \label{ENorlund}
\end{equation}%
The factor $\prod_{l,j,j<l}(r_{l}-r_{j})^{2}$ is responsible for
the coupling between $r_{j}$'s. In order to see the effect of this
coupling, we use the path of integration for each $r_{j}$, which
passes through the stationary point corresponding to a single pair
(line 1 in Fig. 1). This point yields the one-pair energy, as
found by Cooper.

Because of the symmetry between all $r_{j}$'s in the RHS of Eq. (\ref%
{ENorlund}), we may replace $\sum_{j=1}^{N}r_{j}$ by $Nr_{1}$.
Then, $E$ can be rewritten as
\begin{equation}
E=2N\varepsilon _{F_{0}}-N\frac{F_{1}}{F_{0}},  \label{Eratio}
\end{equation}%
where
\begin{eqnarray}
F_{n} &\equiv &\int drr_{1}^{n}\frac{\prod_{l,j,j<l}(r_{l}-r_{j})^{2}}{%
\prod_{j=1}^{N}\prod_{m_{j}=0}^{N_{\Omega
}}(r_{j}+\frac{2m_{j}}{\rho })}\exp \left(
\frac{\sum_{j=1}^{N}r_{j}}{V}\right)   \notag \\
&=&\int drr_{1}^{n}\left[ \prod_{j=1}^{N}\exp \left( \frac{r_{j}}{V}%
-\sum_{m_{j}=0}^{N_{\Omega }}\ln \left( r_{j}+\frac{2m_{j}}{\rho }\right) \right) %
\right] \prod_{l,j,j<l}(r_{l}-r_{j})^{2}. \label{Fn}
\end{eqnarray}

The position of the saddle point is given by the solution of the
following equation:
\begin{equation}
\frac{\partial }{\partial r}f(r)\equiv \frac{\partial }{\partial
r}\left[
\frac{r}{V}-\sum_{m=0}^{N_{\Omega }}\ln \left( r+\frac{2m}{\rho }\right) %
\right] =\frac{1}{V}-\sum_{m=0}^{N_{\Omega
}}\frac{1}{r+\frac{2m}{\rho }}=0. \label{Cooper}
\end{equation}%
We replace sum by the integral that is justified in the
large-sample limit, and find the solution of Eq. (\ref{Cooper}),
$r=\epsilon _{c}$, which gives the binding energy of an isolated
Cooper pair
\begin{equation}
\epsilon _{c}\simeq 2\Omega \frac{\sigma }{1-\sigma },
\label{binding}
\end{equation}%
which actually has already been found in Eq. (\ref{pairener})
through the binomial sum. We now introduce a new variable $t$
defined as $it=r-\epsilon _{c}$. Next, we expand $f(r)$ in a
Taylor series around $r=\epsilon _{c}$ and rewrite it as
\begin{equation}
f(\epsilon _{c}+it)=f(\epsilon _{c})+\varphi (t),  \label{ft}
\end{equation}%
where $\varphi (t)$ reads as
\begin{equation}
\varphi (t)=\sum_{m=2}^{+\infty }\frac{f^{(m)}(\epsilon
_{c})}{m!}(it)^{m}. \label{gexp}
\end{equation}%
Derivatives $f^{(m)}(\epsilon _{c})$ can be easily found from Eqs. (\ref%
{Cooper}) and (\ref{binding}) as
\begin{equation}
f^{(m)}(\epsilon _{c})\simeq (-1)^{m}(m-2)!\frac{\rho }{2}\left( \frac{%
1-\sigma }{2\Omega \sigma }\right) ^{m-1}(1-\sigma ^{m-1}).
\label{fm}
\end{equation}%
Notice that we again have replaced sums by integrals, when deriving Eq. (\ref%
{fm}).

It is seen from Eq. (\ref{fm}) that $f^{(2)}(\epsilon _{c})$ is
positive, since $\sigma <1$. Thus, the direction of the steepest
descent for $\exp (f(r))$ is perpendicular to the real axis in $r$
complex plane, as expected (line 1 in Fig. 1). Hence, we assume
that $t$ changes from $-\infty $ to $+\infty $. Next, we introduce
a rescaled variable $x$ instead of $t$
\begin{equation}
x=t\sqrt{f^{(2)}(\epsilon _{c})}.  \label{rescale}
\end{equation}%
Within new notations, $f$ reads as
\begin{equation}
f\left( \epsilon _{c}+ix/\sqrt{f^{2}(\epsilon _{c})}\right)
=f(\epsilon _{c})+\sum_{m=2}^{\infty }\alpha _{m}x^{m},
\label{fexp}
\end{equation}%
where
\begin{equation}
\alpha _{m}=\frac{(-i)^{m}}{m(m-1)}\frac{1-\sigma ^{m-1}}{1-\sigma
}\delta ^{m-2},  \label{alfa}
\end{equation}%
where $\delta $ is a small dimensionless constant which is
inversely proportional to the square root of the system volume
\begin{equation}
\delta =\frac{1}{\sqrt{\rho \Omega \sigma }}.  \label{gamma}
\end{equation}%
Note that $\alpha _{2}=-1/2$.

The existence of the small constant $\delta $ is crucial in our
derivation, since we are going to calculate the ground state
energy as an asymptotic expansion in powers of $\delta $. We will
however see that $\delta $ is going to be coupled with the number
of pairs, so the resulting expansion of the energy density in the
thermodynamical limit will be in powers of pair density, which is
proportional to $N\delta ^{2}\sim N/\rho $ and is no longer small.
Such an unusual feature is due to the fact that we here deal with
the multidimensional integral, while integration is performed
through the single-pair saddle point.

Next, we rewrite Eq. (\ref{Eratio}) as
\begin{equation}
E=2N\varepsilon _{F_{0}}-N\epsilon _{c}-iN\delta \frac{2\Omega \sigma }{%
1-\sigma }\frac{\Phi _{1}}{\Phi _{0}},  \label{EPhi}
\end{equation}%
where
\begin{equation}
\Phi _{n}=\int dxx_{1}^{n}\left[ \prod_{j=1}^{N}\exp \left(
\sum_{m=2}^{\infty }\alpha _{m}x_{j}^{m}\right) \right]
\prod_{l,j,j<l}(x_{l}-x_{j})^{2},  \label{Phi}
\end{equation}%
where integration for each $x_{j}$ is performed from $-\infty $ to $+\infty $%
.

Let us concentrate on $\Phi _{1}$. It can be integrated by parts
with
respect to $x_{1}$ in the following manner:%
\begin{equation}
\Phi _{1}=\int dx\left[
\prod_{j,l,j<l}(x_{l}-x_{j})^{2}e^{\sum_{m=3}^{\infty }\alpha
_{m}x_{1}^{m}}\right] _{x_{1}}^{\prime }e^{\alpha
_{2}x_{1}^{2}}\prod_{j\neq 1}e^{\sum_{m=2}^{\infty }\alpha
_{m}x_{j}^{m}}. \label{Phi1parts}
\end{equation}%
The derivative in the integrand in the RHS of the above equation
reads
\begin{eqnarray}
&&2\sum_{j_{1}\neq 1}(x_{1}-x_{j_{1}})\prod_{\substack{ l,j,j<l,  \\ %
(j,l)\neq (1,j_{1})}}(x_{l}-x_{j})^{2}e^{\sum_{m=3}^{\infty
}\alpha
_{m}x_{1}^{m}}  \notag \\
&&+\delta \sum_{m=2}^{\infty }a_{m}\delta ^{m-2}x_{1}^{m}\left[
\prod_{l,j,j<l}(x_{l}-x_{j})^{2}e^{\sum_{m=3}^{\infty }\alpha _{m}x_{1}^{m}}%
\right] ,  \label{Phi1deriv}
\end{eqnarray}%
where%
\begin{equation}
a_{m}=\alpha _{m+1}(m+1)\delta
^{-(m-1)}=\frac{(-i)^{m+1}}{m}\frac{1-\sigma ^{m}}{1-\sigma }.
\label{am}
\end{equation}%
The first term in Eq. (\ref{Phi1deriv}) vanishes identically after
we
substitute it into Eq. (\ref{Phi1parts}) and perform integration over all $%
x_{j}$'s, due to the symmetry reasons, which can be easily
verified upon the mutual exchange $x_{1}\longleftrightarrow
x_{j_{1}}$. The second term, after the integration, yields an
expansion of $\Phi _{1}$ through $\Phi _{2}$, $\Phi
_{3}$, etc. with coefficients, which represent higher and higher powers of $%
\delta $
\begin{equation}
\Phi _{1}=\delta \sum_{m=2}^{\infty }a_{m}\delta ^{m-2}\Phi _{m}.
\label{Phi1exp}
\end{equation}

We now focus on the first term in the expansion from the RHS of Eq. (\ref%
{Phi1exp}), which is linear in $\delta $. $\Phi _{2}$ can be
handled in the same manner, as the one used for $\Phi _{1}$.
Namely, we make integration by
parts with respect to $x_{1}$. We then obtain an equation, similar to Eq. (%
\ref{Phi1parts}), except that the expression in square brackets
must be multiplied by $x_{1}$. The derivative of this expression
can be written as
\begin{eqnarray}
&&\prod_{l,j,j<l}(x_{l}-x_{j})^{2}e^{\sum_{m=3}^{\infty }\alpha
_{m}x_{1}^{m}}  \notag \\
&&+2x_{1}\sum_{j_{1}\neq 1}(x_{1}-x_{j_{1}})\prod_{\substack{ l,j,j<l,  \\ %
(j,l)\neq (1,j_{1})}}(x_{l}-x_{j})^{2}e^{\sum_{m=3}^{\infty
}\alpha
_{m}x_{1}^{m}}  \notag \\
&&+\delta \sum_{m=2}^{\infty }a_{m}\delta ^{m-2}x_{1}^{m+1}
\prod_{l,j,j<l}(x_{l}-x_{j})^{2}e^{\sum_{m=3}^{\infty }\alpha
_{m}x_{1}^{m}}.  \label{Phi2deriv}
\end{eqnarray}%
After integration, the first term in the RHS of the above equation gives $%
\Phi _{0}$. The second term is the sum of $N-1$ terms. By
performing a mutual exchange $x_{1}\longleftrightarrow x_{j_{1}}$,
it is easy to see that each of these terms also gives $\Phi _{0}$
after the integration. The third term leads to the linear
combination of $\Phi _{m}$'s with $m$ starting from $3$. As a
result, $\Phi _{2}$ reads
\begin{equation}
\Phi _{2}=N\Phi _{0}+\delta \sum_{m=2}^{\infty }a_{m}\delta
^{m-2}\Phi _{m+1}.  \label{Phi2exp}
\end{equation}

At this stage, we substitute Eq. (\ref{Phi2exp}) to the expression
of $\Phi _{1}$, as given by Eq. (\ref{Phi1exp}), and obtain
\begin{eqnarray}
\Phi _{1} &=&\delta N\Phi _{0}a_{2}+\delta ^{2}\sum_{m=2}^{\infty
}a_{m+1}^{(1)}\delta ^{m-2}\Phi _{m+1}  \notag \\
&=&\delta N\Phi _{0}a_{2}+O(\delta ^{2}),  \label{Phi1exp2}
\end{eqnarray}%
where $a_{m+1}^{(1)}$ is a "new" expansion coefficient, given by
\begin{equation}
a_{m+1}^{(1)}=a_{m+1}+a_{2}a_{m}.  \label{a1}
\end{equation}%
Eq. (\ref{Phi1exp2}) provides $\Phi _{1}/\Phi _{0}$ in lowest order in $%
\delta $. It is then substituted to the expression for the energy,
given by Eq. (\ref{EPhi}). Together with Eq. (\ref{am}) for $m=2$
and the definition of $\delta $ it leads to
\begin{equation}
E=E_{N}+\frac{N}{\rho}\frac{1+\sigma}{1-\sigma}+O\left( \rho
^{-3/2}\right) , \label{Ebasic}
\end{equation}%
where $E_{N}$ is given by Eq. (1). The second term in the RHS of
Eq. (\ref{Ebasic}) is underextensive, so that it can be dropped.
Of course, one has to keep in mind that the third term, $%
O\left( \rho ^{-3/2}\right) $, is not necessarily small due to the
coupling with $N$.

Thus, we have calculated the correction to the energy of $N$
noninteracting pairs in the lowest order in $\delta $, which gives
nonzero contribution,
i.e., in $\delta ^{2}$. This correction turns out to be proportional to $%
N(N\delta ^{2})$ (due to the coupling with $N$) that is extensive.
Moreover, this first correction already gives the total energy
within all relevant (extensive) terms, as evident from the results
of Section III. The task is now to prove that the third term in
the RHS of Eq. (\ref{Ebasic}), produces \textit{underextensive}
contribution only. This problem is addressed in the Appendix B,
where we show, through rather tedious calculations, that two next
terms in $\delta $ are indeed underextensive. We stress that we
are unable to present a complete proof involving all terms of the
expansion. The underextensivity we have revealed follows from
"magic rules" for coefficients $a_{m}$, which are of course linked
to the constant density of states. Thus, the method based on
calculation of the binomial sum, which enabled us to derive the
expression for the ground state energy directly within all
extensive terms in a simple way (nonperturbatively in $\delta $),
appears to be much more efficient than the method of the present
Section.

We also wish to stress that the method of saddle-point evaluation of N\"{o}%
rlund-Rice integral through the single-pair solution is quite
similar to the
method of analytical solution of Richardson equations presented in Ref. [%
\cite{We}] and advanced further in Ref. [\cite{EPJB}]. In Ref.
[\cite{We}], all Richardson energy-like quantities $R$'s have been
expanded around a single-pair solution in seria involving small
dimensionless parameter analogous to $\delta $. By keeping only
the lowest-order contribution, Eq. (1) has been derived. The
lowest-order approximation a priori makes this method restricted
to the dilute regime of pairs only, when their density is small.
However, in Ref. [\cite{EPJB}] it was demonstrated that next term
in the small dimensionless parameter vanishes identically due to
the first "magic cancellation", which implies that the
dilute-limit result of Ref. [ \cite{We}] is likely to be
universal. We have arrived to the similar conclusion within our
framework, but we have also revealed the second "magic
cancellation" which exists for the next-order term (see the
Appendix B).

The method of integration through the single-pair saddle point, as
presented in this Section, has some disadvantages (in addition to
its obvious technical complexity compared to the method based on
manipulations with binomial sums). Namely, it relies on replacing
sums by integrals, as we did to derive Eq. (\ref{binding}). More
rigorously, sums of this kind should be expressed through $\Gamma
$-functions. By keeping a leading order term in the expansion of
these $\Gamma $-functions in $1/\rho $, one obtains the
approximation, used here. Dropping all the remaining terms
introduces errors of the order of $\delta ^{2}\sim 1/\rho $ in
many places. Fortunately, this approximation does not lead to any
pathologies for extensive quantities, but it produces artificial
underextensive terms (not considered here). In addition, the
effect of the reduction of a pair binding energy in a many-pair
configuration\cite{We} compared to the isolated pair is clearly of
discrete origin, while, within this approach, it is recovered
through the continuous approximation, i.e., in a very indirect
way.

\section{Conclusions}

Within the exact mapping of Richardson equations to the system of
interacting charges in the plane, we suggested to switch from the
"energy" of this system to the "probability" for charges to occupy
certain states in configurational space at the effective
temperature given by the interaction constant. The effective
temperature thus goes to zero when system volume goes to infinity.
We introduced a "partition function", from which the ground state
energy of the initial quantum-mechanical many-body problem can be
found. This approach leads to the emergency of quite rich
mathematical structure. The "partition function" has a form of a
multidimensional integral, similar to Selberg integrals. For the
model with constant density of energy states, the structure of the
integrand implies that it can be also considered as an integral of
N\"{o}rlund-Rice type. The most efficient way to evaluate it is by
transforming the integral into a binomial sum, where the coupling
between variables is due to the determinant of the Vandermonde
matrix. Using properties of this matrix, we managed to evaluate
the "partition function" exactly in a rather simple way and to
find the ground state energy, which is described by a single
expression all over from the dilute to the dense regime of pairs.
This expression does coincides with the mean-field BCS result.

We also provided a qualitative understanding of the obtained
result in terms of "probabilities". Namely, the "probability" of
the system of $N$ pairs, feeling each other through the Pauli
exclusion principle, to be in a certain state can be represented
in a factorized form as a linear combination of terms. Each of
them is given by the product of "probabilities" for $N$ single
pairs, but every single pair is placed into its own environment,
which is identical to the initial one, except that two bands of
the one-electronic energy states - both from the bottom and from
the top of energy interval - are absent due to the Pauli exclusion
principle. Moreover, we find that although these environments are
different, the sum of energies of the single pairs is the same for
all terms of factorized "probability".

Finally, we presented another method for evaluation of the
N\"{o}rlund-Rice integral by integration through the saddle point
corresponding to a single pair. This method turns out to be much
less efficient; it exactly corresponds to the dilute-limit
approach for the solution of Richardson equation proposed very
recently in Refs. \cite{We,EPJB}. Nevertheless, we managed to
calculate several initial terms of the expansion of energy in
pairs density.

The suggested method is rather general and we believe that it can
be applied to other integrable pairing Hamiltonians, which have
electrostatic analogies\cite{Amico,Ibanez}, as well as to
situations with nonconstant density of energy states.

\begin{acknowledgments}
The author acknowledges numerous discussions with Monique
Combescot, as well as helpful comments by A. O. Sboychakov. This
work is supported by the Dynasty Foundation, the Russian
Foundation for Basic Research (project no. 09-02-00248), and, in
parts, by the French Ministry of Education.
\end{acknowledgments}

\appendix

\section{Electrons-holes duality}

We introduce creation operators for holes as
$b_{\mathbf{k}\uparrow }^{\dagger }=a_{\mathbf{k}\uparrow }$ and
$b_{\mathbf{k}\downarrow }^{\dagger }=a_{\mathbf{k}\downarrow }$
with corresponding rules for destruction operators. These holes
should not be confused with holes in the Fermi sea of normal
electrons, as appear in BCS theory. By using commutation relations
for fermionic operators, it is rather easy to rewrite the initial
Hamiltonian in terms of
holes as%
\begin{equation}
H=-V\sum_{\mathbf{k}}1+2\sum_{\mathbf{k}}\varepsilon _{\mathbf{k}}-\sum_{%
\mathbf{k}}\left( \varepsilon _{\mathbf{k}}-V\right) \left( b_{\mathbf{k}%
\uparrow }^{\dagger }b_{\mathbf{k}\uparrow
}+b_{\mathbf{k}\downarrow
}^{\dagger }b_{\mathbf{k}\downarrow }\right) -V\sum_{\mathbf{k},\mathbf{k}%
^{\prime }}b_{\mathbf{k}^{\prime }\uparrow }^{\dagger }b_{-\mathbf{k}%
^{\prime }\downarrow }^{\dagger }b_{-\mathbf{k}\downarrow }b_{\mathbf{k}%
\uparrow }.  \label{BCSholes}
\end{equation}%
The first two terms of the RHS of Eq. (\ref{BCSholes}) are
scalars. Moreover, the first term,\ $V\sum_{\mathbf{k}}1$, is
inderextensive and can be dropped. The fourth term in the RHS of
Eq. (\ref{BCSholes}) is fully
identical to the interaction potential in terms of electrons, given by Eq. (%
\ref{BCSpot}). To better understand the role of the third term, we
introduce $\xi _{\mathbf{k}}^{^{\prime }}$ defined as $\xi
_{\mathbf{k}}^{^{\prime }}=$ $\varepsilon _{F_{0}}+\Omega
-\varepsilon _{\mathbf{k}}$. In the case of a constant density of
states, $\xi _{\mathbf{k}}^{^{\prime }}$ is $0$, $1/\rho $,
$2/\rho $, ..., $\Omega $: $\xi _{\mathbf{k}}^{^{\prime }}$ just
counts
states starting from the top of potential layer towards its bottom. Hence, $%
-(\varepsilon _{\mathbf{k}}-V$) can be rewritten as $\xi _{\mathbf{k}%
}^{^{\prime }}-(\varepsilon _{F_{0}}+\Omega -V)$. A similar term
in the
Hamiltonian for electrons, given by Eq. (\ref{H0}), contains factor $\xi _{%
\mathbf{k}}+\varepsilon _{F_{0}}$. Thus, we clearly see a duality
between electrons and holes: The ground state of $N$ pairs is
equivalent to the ground state of $N_{\Omega }-N$ holes, while the
energy of the latter is equal to the energy of $N_{\Omega }-N$
pairs, with $\varepsilon _{F_{0}}$ changed into $-(\varepsilon
_{F_{0}}+\Omega -V)$, plus the kinetic energy of
the potential layer completely filled, given by $2\sum_{\mathbf{k}%
}\varepsilon _{\mathbf{k}}$. This means that we can avoid
considering configurations with $N>N_{\Omega }/2$ by switching to
holes, Cooper pairs in this case being constructed out of them.
Moreover, one can directly check that the expression of the ground
state energy given by Eq. (1), which we are going to prove,
satisfies, within underextensive terms, the above duality
criterion. Therefore, it is sufficient to consider the case
$N<N_{\Omega }/2$ only. The existence of the duality between
electrons and holes in the energy spectrum has already been noted
in Ref. [\cite{We,MWO}], although its origin, at the level of the
Hamiltonian, stayed unclear.

\section{Cancellation of higher-order terms in pairs density}

We focus on $\Phi _{3}$ and rewrite it through the integration by
parts, as has been done for $\Phi _{1}$ and $\Phi _{2}$
\begin{equation}
\Phi _{3}=2N\Phi _{1}+\delta \sum_{m=2}^{\infty }a_{m}\delta
^{m-2}\Phi _{m+2}.  \label{Phi3exp}
\end{equation}%
Next, we substitute Eq. (\ref{Phi3exp}) to the first line of Eq. (\ref%
{Phi1exp2}) and get the following expression of $\Phi _{1}/\Phi
_{0}$
\begin{equation}
\frac{\Phi _{1}}{\Phi _{0}}=\delta Na_{2}\frac{1}{1-2\delta ^{2}Na_{3}^{(1)}}%
\left( 1+\frac{\delta ^{2}}{Na_{2}}\sum_{m=3}^{\infty
}a_{m+1}^{(2)}\delta ^{m-3}\frac{\Phi _{m+1}}{\Phi _{0}}\right) ,
\label{Phi1exp3}
\end{equation}%
where
\begin{equation}
a_{m+1}^{(2)}=a_{m+1}^{(1)}+a_{3}^{(1)}a_{m-1}.  \label{a2}
\end{equation}

We now consider the first term of the sum in the RHS of Eq. (\ref{Phi1exp3}%
). Again integrating by parts, we obtain
\begin{equation}
\Phi _{4}\simeq 2N\Phi _{2}+N\Phi _{1,1}+\delta \sum_{m=4}^{\infty
}a_{m-2}\delta ^{m-4}\Phi _{m+1},  \label{Phi4exp}
\end{equation}%
while $\Phi _{m,n}$ is given by Eq. (\ref{Phi}) with $x_{1}^{n}$
replaced by $x_{1}^{m}x_{2}^{n}$. $\Phi _{1,1}$ is evaluated as
\begin{equation}
\Phi _{1,1}=-\Phi _{0}+\delta \sum_{m=1}^{\infty }a_{m+1}\delta
^{m-1}\Phi _{1,m+1}.  \label{Phi11exp}
\end{equation}%
Now we substitute Eq. (\ref{Phi11exp}) for $\Phi _{1,1}$ and Eq. (\ref%
{Phi2exp}) for $\Phi _{2}$ to Eq. (\ref{Phi4exp}) for $\Phi _{4}$,
and the
resulting equation for $\Phi _{4}$ - to Eq. (\ref{Phi1exp3}) for $\Phi _{1}$%
. The obtained expression of $\Phi _{1}$ can be written as a sum
of two
contributions%
\begin{equation}
\Phi _{1}=G_{1}+G_{2},  \label{Phi1sum}
\end{equation}%
where
\begin{equation}
G_{1}=\delta N\frac{\Phi _{0}a_{2}}{1-2a_{3}^{(1)}\delta ^{2}N}\left( 1+%
\frac{2a_{4}^{(2)}}{a_{2}}\delta ^{2}N\right),  \label{G1}
\end{equation}%
\begin{equation}
G_{2}=\frac{\delta ^{4}}{1-2a_{3}^{(1)}\delta ^{2}N}\left(
\sum_{m=4}^{\infty }a_{m+1}^{(3)}\delta ^{m-4}\Phi
_{m+1}+a_{4}^{(2)}N\sum_{m=2}^{\infty }a_{m}\delta ^{m-2}(2\Phi
_{m+1}+\Phi _{1,m})\right),  \label{G2}
\end{equation}%
where
\begin{equation}
a_{m+1}^{(3)}=a_{m+1}^{(2)}+a_{4}^{(2)}a_{m-2}.  \label{a3}
\end{equation}

It is can be checked that
\begin{equation}
a_{4}^{(2)}+a_{3}^{(1)}a_{2}=0  \label{magic1}
\end{equation}%
for any $\sigma $. Therefore, due to the exact cancellation
between the numerator and denominator, Eq. (\ref{G1}) for $G_{1}$
is reduced to
\begin{equation}
G_{1}=(\delta N)\Phi _{0}a_{2}.  \label{G1fin}
\end{equation}%
This means that terms proportional to $N\delta (N\delta
^{2})=N^{2}\delta ^{3}$, which are present in $G_{1}$ and not in
$G_{2}$, are absent in the expansion of $\Phi _{1}/\Phi _{0}$ in
powers of $\delta $. As clearly seen from Eq. (\ref{EPhi}), these
are exactly the terms, which give contribution of the order of
$N(N/\rho )^{2}\sim N(N\delta ^{2})^{2}$ to the energy, or
equivalently, terms in square of pairs density ($\sim (N/\rho
)^{2}$) to the energy density. Eq. (\ref{magic1}) constitutes
first "magic cancellation" rule and it is fully equivalent to Eq.
(68) of Ref. [\cite{EPJB}].

Let us now make some comments. In general, at each step of our
procedure, we express $\Phi _{m}$ through the linear combination
of all possible terms of
the form $\Phi _{m_{1},m_{2},...}$, with the sum of nonnegative integer $m$%
's equal to $m-2$ and prefactors, independent on $\delta $, plus
expansion which involves $\delta ^{n}\Phi _{m+n}$, $n$ starting
from 1. We then have to proceed with all these "new" $\Phi
_{m_{1},m_{2},...}$'s to lower the sum of $m$'s by 2 at each
iteration and finally to bring them either to $\Phi _{0}$ or to
$\Phi _{1}$, depending on a parity of $m$. And so on. Of course,
this recursive tree-like procedure becomes more and more tedious
when $m$ is increasing. That is why we were able to trace only few
first terms of the expansion.

Now we provide a sketch for the derivation of the next-order
correction to the energy, which will be of the order of $N(N/\rho
)^{3}$. For that, we first have to consider $\Phi _{5}$. We
express $\Phi _{5}$ through $\Phi _{3}$ (given by Eq.
(\ref{Phi3exp})), $\Phi _{1,2}$, and $\Phi _{n}$ for $n$'s
starting from $6$. $\Phi _{1,2}$ is expressed through $\Phi _{1}$
and $\Phi _{2,n}$, with $n$ starting from 2. Then, we perform
similar manipulations
for $\Phi _{6}$. Next, we substitute all these quantities to Eqs. (\ref%
{Phi1sum})-(\ref{G2}) for $\Phi _{1}$ and obtain more
sophisticated fraction than the one in the RHS of Eq.
(\ref{Phi1exp3}) which now includes higher powers of $\delta $.

By expanding this "new" fraction in $\delta$, we finally ensure
that terms of $\Phi _{1}/\Phi _{0}$ in $N\delta (N\delta
^{2})^{2}=N^{3}\delta ^{5}$ vanish due to the second "magic
cancellation" rule, given by
\begin{equation}
5a_{6}^{(4)}+10a_{5}^{(3)}a_{2}+a_{4}^{(2)}(4a_{3}+9a_{2}^{2})=0,
\label{magic2}
\end{equation}%
where
\begin{equation}
a_{m+1}^{(4)}=a_{m+1}^{(3)}+a_{5}^{(3)}a_{m-3}.  \label{a4}
\end{equation}%
Again, Eq. (\ref{magic2}) is fulfilled for any $\sigma $. Then, it
follows from Eq. (\ref{EPhi}) that terms in $N(N/\rho )^{3}\sim
N(N\delta ^{2})^{3}$ vanish in the expression of the
quantum-mechanical energy.

\end{document}